\documentclass[onecolumn]{IEEEtran}
\usepackage{amsmath}
\usepackage{theorem}
\usepackage{amsmath}
\usepackage{mathrsfs}
\usepackage{accents}
\usepackage{url}
\usepackage{algorithm}
\usepackage{algorithmic}
\usepackage{epsfig}
\usepackage{amsfonts}
\usepackage{amssymb}
\usepackage{mathrsfs}
\usepackage{euscript}
\usepackage{graphicx}
\usepackage{multirow}
\usepackage{cite}
\usepackage{epsfig}
\usepackage{graphics}
\usepackage[protrusion=true,expansion=true]{microtype} 
\usepackage{graphicx} 
\usepackage{wrapfig} 
\usepackage{epstopdf}
\usepackage{amsmath}
\usepackage{array}
\usepackage{booktabs}
\usepackage{tabularx}
\usepackage{tabulary}
\newtheorem{theorem}{Theorem}
\newtheorem{lemma}{Lemma}

\usepackage{algorithm}
\usepackage{algorithmic}
\usepackage{setspace}
\usepackage[nodayofweek,level]{datetime}
\newcommand{\mydate}{\formatdate{20}{12}{2018}}
\newcommand\blfootnote[1]{%
	\begingroup
	\renewcommand\thefootnote{}\footnote{#1}%
	\endgroup
}
\begin{document}
	
	\begin{titlepage}

		\begin{tabular}{l r}
			
			\includegraphics[scale=0.3]{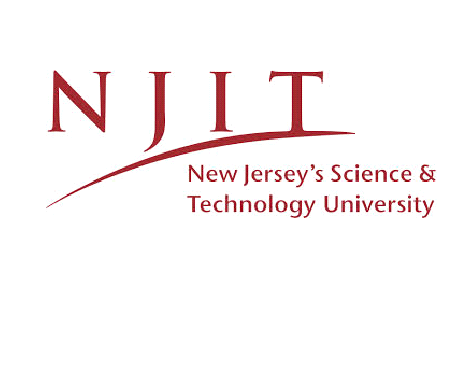} \hspace{10cm} & \includegraphics[scale=0.3]{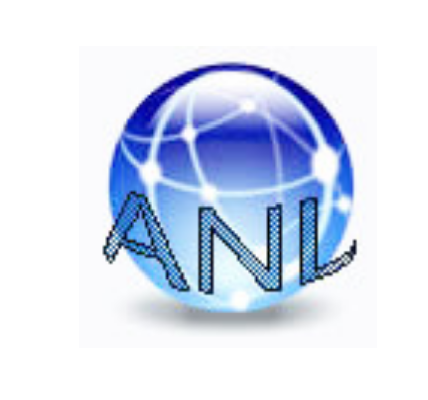}
			
		\end{tabular}
		
		\begin{center}
			\vspace{10mm}
			
			\textsc{\LARGE NOMA Aided Narrowband IoT for Machine Type Communications with User Clustering}\\[4cm]
			
			{\Large \textsc{Ali Shahini}}\\ 
			{\Large \textsc{Nirwan Ansari}}\\ 
			[4cm]
			
			{}
			{\textsc{TR-ANL-2018-002}\\
				\large \mydate} \\[4cm]
			
			{\textsc{Advanced Networking Laboratory}}\\
			{\textsc{Department of Electrical and Computer Engineering}}\\
			{\textsc{New Jersey Institute of Technology}}\\[1.5cm]
			\vfill
			
		\end{center}
		
	\end{titlepage}
\onecolumn

\begin{abstract}
To support Machine Type Communications (MTC) in next generation mobile networks, NarrowBand-IoT (NB-IoT) has been released by the Third Generation Partnership Project (3GPP) as a promising solution to provide extended coverage and low energy consumption for low cost MTC devices. However, the existing Orthogonal Multiple Access (OMA) scheme in NB-IoT cannot provide connectivity for a massive number of MTC devices. In parallel with the development of NB-IoT, Non-Orthogonal Multiple Access (NOMA), introduced for the fifth generation wireless networks, is deemed to significantly improve the network capacity by providing massive connectivity through sharing the same spectral resources. To leverage NOMA in the context of NB-IoT, we propose a power domain NOMA scheme with user clustering for an NB-IoT system. In particular, the MTC devices are assigned to different ranks within the NOMA clusters where they transmit over the same frequency resources. Then, we formulate an optimization problem to maximize the total throughput of the network by optimizing the resource allocation of MTC devices and NOMA clustering while satisfying the transmission power and quality of service requirements. We prove the NP-hardness of the proposed optimization problem. We further design an efficient heuristic algorithm to solve the proposed optimization problem by jointly optimizing NOMA clustering and resource allocation of MTC devices. Furthermore, we prove that the reduced optimization problem of power control is a convex optimization task. Simulation results are presented to demonstrate the efficiency of the proposed scheme.
\end{abstract}

\blfootnote{The authors are with Advanced Networking Laboratory, the Helen and John C. Hartmann Department of Electrical and Computer Engineering, New Jersey Institute of Technology, Newark, NJ 07102.\protect~E-mail: {ali.shahini, nirwan.ansari}@njit.edu. This work is currently under review in IEEE Internet of Things Journal.}

\section{Introduction}\label{sec:Introduction}
\IEEEPARstart{I}{nternet} of Things (IoT) is a world wide network of interconnected entities and is anticipated to grow in coming years with the projection of connecting as many as billions of devices with an average of 6-7 devices per person by 2020~\cite{XilongIoT,CiscoWhite}. There are three typical usage scenarios for fifth generation (5G) mobile network services, including  enhanced  mobile  broadband  (eMBB), massive machine type communications (mMTC) and ultra-reliable and low-latency communications (URLLC)~\cite{5Gembburllcmmtc}. Different from eMBB, mMTC and URLLC mainly target services of IoT and are considered as two types of Machine Type Communications (MTC) characterized by the International Telecommunications Union (ITU). mMTC and URLLC devices as two important enablers of IoT have different characteristics. mMTC requires connectivity of a massive number of active low-power devices in co-existence in one cell, and these devices transmit small packets with relaxed latency requirements in the order of seconds or hours~\cite{mmtcReq}. Unlike mMTC, ultra reliable data transmissions is essential for URLLC devices along with low latency requirements as they are used for critical applications~\cite{5Gembburllcmmtc}.

To support MTC for next generation mobile networks, a new technology called Narrow-band Internet of Things (NB-IoT) has recently been standardized by  the Third Generation Partnership Project (3GPP) in its Release 13 \cite{3gpprel13}. In particular, NB-IoT provides energy efficient communications for low power MTC devices on a narrow bandwidth of 180 kHz for both downlink and uplink \cite{NB-IoTMag}. In order to provide better granularity and higher utilization, the unit of resource scheduling in the NB-IoT uplink is sub-carrier instead of Physical Resource Block (PRB). In fact, the NB-IoT uplink has sub-carrier spacing of 3.75 kHz, i.e., the minimum transmission bandwidth for a device, whereas the downlink retains the Long Term Evolution (LTE) downlink transmission structure with 15 kHz sub-carrier spacing \cite{NB-IoTSubCarrier}. NB-IoT can provide data rates of nearly 250 kbps in downlink and 20 kbps in uplink transmissions with the possibility to aggregate multiple sub-carriers to reach the downlink speed \cite{NBIOT-Shirvani,NBiot3class}.
The target of NB-IoT is to prolong the battery lifetime to reach 10 years and provide massive connectivity of devices \cite{NB-IoTMag}. However, the main challenge of providing connectivity to a massive number of MTC devices in 5G networks cannot be addressed by existing NB-IoT technologies.

Currently, NB-IoT exploits an orthogonal multiple access (OMA) scheme over a bandwidth of 180 kHz where each sub-carrier cannot be occupied by more than one user. Thus, the OMA scheme in NB-IoT fails to cope with the massive increase in the number of connected MTC devices. Hence, to support connectivity to a massive number of MTC devices with the limited number of sub-carriers in one PRB, a promising solution is to adopt power-domain Non-Orthogonal Multiple Access (NOMA) scheme \cite{Connectivity11,NOMAmagazine}. In contrast with OMA methods, NOMA supports massive connectivity by allocating multiple MTC devices to share each sub-carrier. In other words, multiple MTC devices can transmit over the same frequency resources, thus resulting in a significant increase in the network connectivity. In the power domain NOMA scenario, different power level strategy is considered to decode the differentiated messages sequentially at the receiver side \cite{JSACNoma}. In fact, the Successive Interference Cancellation (SIC) \cite{NOMAvtcSIC} scheme is exploited at the receiver side to extract the transmitted messages. Thus, NOMA can help NB-IoT systems to meet their demands of massive connectivities, and high spectral-energy efficiency.

\subsection{Contributions}

While there are several research activities that investigate NOMA techniques for 5G networks, none, to our best knowledge, has leveraged the advantages of NOMA in  the context of NB-IoT with user clustering of different users with various quality of service (QoS) requirements. To  this  end, we aim to address the aforementioned issue by proposing a general system model focusing on two emerging technologies of NOMA and NB-IoT. In fact, we propose a novel NOMA based NB-IoT model to maximize the total throughput of an NB-IoT network by increasing the number of connected devices through optimal clustering of MTC devices and optimizing the resource allocation. In particular, MTC devices are grouped into different NOMA clusters and share the same frequency resources among the cluster members. Considering the intra-cell interferences, transmission power and QoS requirements, the MTC devices are ranked in each NOMA cluster. The goal is to maximize the total uplink transmission rate of MTC devices by optimizing NOMA clustering and resource allocation of MTC devices. The main contributions of the paper include:
\begin{itemize}
	\item We propose a NOMA clustering method for MTC devices in an NB-IoT system. In particular, MTC devices are classified into different NOMA clusters and the same frequency resources are shared among the cluster members. Considering the intra-cell interferences, transmission power and QoS requirements, the MTC devices are ranked in each NOMA cluster. Therefore, spectral resources are allocated to the NOMA clusters based on the requirements of NOMA cluster members.	
	\item We formulate a NOMA based optimization problem to maximize the total sum rate of uplink transmission in an NB-IoT system by optimizing the resource allocation of MTC devices and NOMA clustering while satisfying the transmission power and quality of service requirements. We further prove the NP-hardness of the proposed optimization problem.
	\item We propose an efficient heuristic algorithm to solve the optimization problem by jointly optimizing NOMA clustering and resource allocation of MTC devices. Furthermore, we prove that the reduced optimization problem of power control is a convex optimization task by introducing variable transformations.
	\item We evaluate the performance of our proposal and the heuristic algorithm via simulations to demonstrate the benefits of NOMA in increasing the total throughput of MTC devices in an NB-IoT system.
\end{itemize}

\subsection{Related Works} 

In this section, related works including NB-IoT, NOMA, and resource allocation are discussed. In the past few years, several works investigated the major challenges of NB-IoT and researchers came up with different algorithms and models. Recently, Yang \textit{et al.} \cite{NBIOTSmallCellAssisted} investigated the small-cell assisted traffic offloading for NB-IoT systems and formulated a joint traffic scheduling and power allocation problem to minimize the total power consumption. Oh and Shin \cite{NBIOTSmallDataLetter} proposed an efficient small data transmission scheme for NB-IoT in which devices that are in an idle state can transmit a small data packet without the radio resource control connection. Malik \textit{et al.} \cite{NBIOTRadioResource} investigated radio resource management in NB-IoT systems by proposing an interference aware resource allocation for the rate maximization problem.

Al-Imari \textit{et al.} \cite{TafazolliNOMA} proposed a NOMA scheme for uplink data transmission that allows multiple users to share the same sub-carrier without any coding/spreading redundancy.  Mostafa \textit{et al.} \cite{Connectivity} studied the connectivity maximization for the application of NOMA in NB-IoT, where only two users can share the same sub-carrier. Kiani and Ansari \cite{NOMAabbas} proposed an edge computing aware NOMA technique in which MEC users' uplink energy consumption is minimized via an optimization framework. Wu \textit{et al.} \cite{WuIoTSE} investigated the spectral efficiency maximization problem for wireless powered NOMA IoT networks. Shahini \textit{et al.} \cite{ShahiniIoT} proposed the energy efficiency maximization problem for cognitive radio (CR) based IoT networks by taking into consideration of user buffer occupancy and data rate fairness. Qian \textit{et al.} \cite{SICNOMAawareMEC} proposed an optimal SIC ordering to minimize the maximum task execution latency across devices for MEC-aware NOMA NB-IoT network. Zhai \textit{et al.} \cite{ZhaiIoT} proposed a joint user scheduling and power allocation for NOMA based wireless networks with massive IoT devices. Xu and Darwazeh \cite{DoubleConnectedNBIoT} proposed a compressed signal waveform solution, termed fast-orthogonal frequency division multiplexing (Fast-OFDM), to potentially double the number of connected devices.

Several works have investigated NOMA for 5G networks, but none has looked into employing NOMA clustering for NB-IoT users with various QoS requirements. Therefore, we propose a novel NOMA based NB-IoT model to maximize the total throughput of the network by optimizing both NOMA clustering and the resource allocation of MTC devices in an NB-IoT system.

The remainder of this paper is organized as follows. In Section II, we describe the system model including NOMA clustering and QoS constraints. In Section III, we formulate the framework of the throughput maximization problem. In Section IV, we detail the proposed algorithm. In Section V, numerical results and simulations are presented. Finally, we conclude the paper in Section VI.

\section{System Model}\label{sec:System Model}
\begin{figure} [t]
	\centering
	\includegraphics[width=12cm,height=7cm]{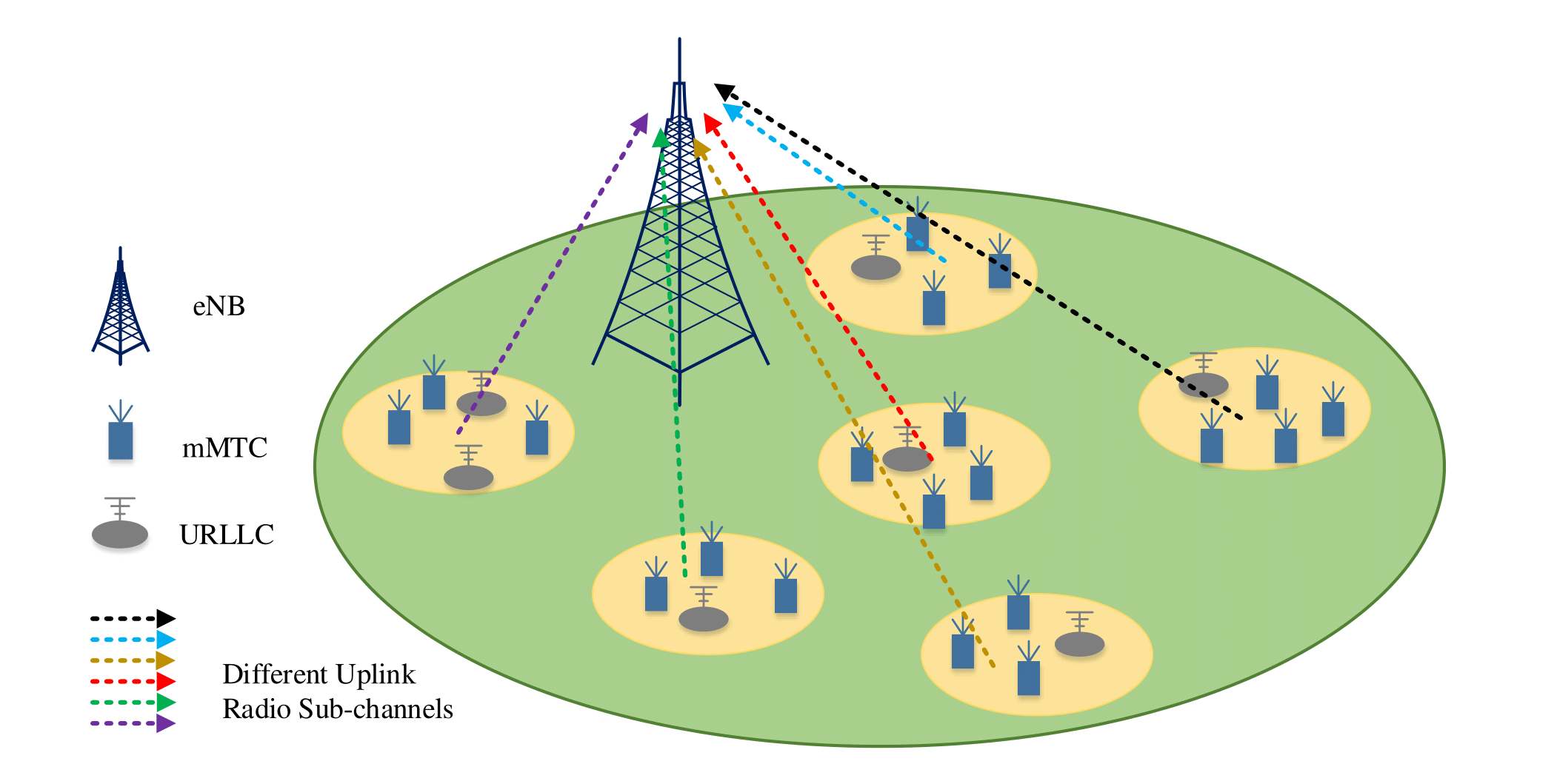}
	\caption{The NOMA clusters include mMTC and URLLC devices, where the allocated sub-channels to each NOMA cluster are shared by the MTC devices.}
	\label{fig:sysmodel}
\end{figure}
We consider a single-cell scenario with one eNB, which supports MTC based on NB-IoT standard \cite{3gpprel13}. We assume there is no inter-cell interference from other neighboring cells. Denote $\mathcal{U} = \left\{ {1,...,U} \right\}$ and $\mathcal{M} = \left\{ {1,...,M} \right\}$ as the sets of mMTC and URLLC devices, respectively. Active URLLC and mMTC devices share one physical resource block (PRB) for uplink data transmission in one transmission time interval (TTI). The available bandwidth of one PRB is assumed to be divided into a set of sub-channel frequencies $\mathcal{S} = \left\{ {1,...,S} \right\}$ and the bandwidth of each sub-channel is $W$.  In fact, the system bandwidth can be equally divided into either 48 or 12 sub-carriers in NB-IoT systems. In particular, the sub-carrier spacing of 3.75 kHz can only be supported for uplink transmissions \cite{NBiot3class}. Therefore, we consider one PRB with 48 sub-carriers of 3.75 kHz for the uplink data transmissions.

We propose a power-domain NOMA scheme by clustering mMTC and URLLC devices in an NB-IoT network as shown in Fig.~\ref{fig:sysmodel}. According to the NOMA scheme, the mMTC and URLLC devices share each sub-carrier (sub-channel), and transmit data in a non-orthogonal manner, i.e., more than one user can share the same sub-channel. Therefore, the devices are divided into different groups, called the NOMA clusters. Denote $\mathcal{C} = \left\{ {1,...,C} \right\}$ as the set of NOMA clusters, and $\gamma^{s,c}$ as the binary variable to assign sub-channel $s \in \mathcal{S}$ to NOMA cluster $c \in \mathcal{C}$. Hence,  ${\gamma ^{s,c}}=1$ if sub-channel $s$ is allocated to the $c^{th}$ NOMA cluster, and ${\gamma ^{s,c}}=0$ otherwise. The URLLC and mMTC devices transmit their messages on the same sub-channel with transmission powers of $p_u$ and $p_m$, respectively. Thus, a combined message from URLLC and mMTC devices with additive noise $N_0$ is received at the eNB. In order to successfully decode messages from the combined received message, the eNB employs the SIC scheme. Thus, the users need to be ordered in each cluster according to the SIC method.

Define the set of the order (ranks) in each cluster as $\mathcal{K} = \left\{ {1,...,{k_{\max }}} \right\}$, where $k_{\max }$ specifies the maximum number of users that are allowed to be in one cluster and consequently share the allocated sub-channels. Note that we assume $C \times {k_{\max }}$ should be greater than the total number of the devices. According to the principles of SIC \cite{NOMAvtcSIC}, the $k^{th}$ user's message in each cluster is decoded before the other users with higher orders. Therefore, the users with higher ranks ($\left\{ {k+1,k+2,...} \right\}$) in each cluster introduce interference to the $k^{th}$ user. In other words, the user with the highest rank in each cluster does not experience interference from other users and the first user receives interference from other users with higher ranks ($k = 2,...,{k_{\max }}$). Note that URLLC devices have higher data rate requirements as compared to mMTC devices. Thus, the transmission power of URLLC devices are higher than that of mMTCs. Therefore, in each cluster, the URLLC devices are required to have lower ranks as compared to mMTC devices. In fact, the SIC decoder at the eNB starts decoding with URLLCs, and consequently the mMTC devices are not affected by high interference from URLLCs. For the ease of reading, frequently used notations and terminologies are summarized in Table \ref{tableSymbols}.

\subsection{Quality of Service Constraints}

	\begin{table}  \label{tableSymbols}	\caption{List of Symbol Notations and Description}
		\centering
		\begin{tabular}{ |l|p{12.5cm}| }
			\hline
			\textbf{Symbols} & \textbf{Descriptions}\\\hline \hline
			$\mathcal{U}$ ($\mathcal{M}$) & The set of URLLC users (the set of mMTC users) \\\hline
			$\mathcal{S}$ & The set of sub-channels in an NB-IoT system\\\hline
			$\mathcal{C}$ ($\mathcal{K}$) & The set of NOMA clusters (the set of orders in each NOMA cluster)\\\hline
			$k_{max}$ & The maximum number of users in one NOMA cluster\\\hline
			$\gamma ^{s,c}$ & The binary indicator whether to allocate the $s^{th}$ sub-channel to the $c^{th}$ cluster\\\hline
			$p^{s}_{u}$ &  The transmission power of the $u^{th}$ URLLC device over the $s^{th}$ channel\\\hline
			$p^{s}_{m}$ &  The transmission power of the $m^{th}$ mMTC device over the $s^{th}$ channel\\\hline
			$N_0$ & The Additive White Gaussian Noise\\\hline
			$\alpha_m^{c,k}$  & The binary indicator whether to assign the $m^{th}$ mMTC to the $k^{th}$ order of cluster $c$ \\\hline
			$\beta_u^{c,k}$  & The binary indicator whether to assign the $u^{th}$ URLLC to the $k^{th}$ order of cluster $c$ \\\hline
			$R_m$ & The total transmission rate of the $m^{th}$ mMTC device \\\hline
			$R_u$ & The total transmission rate of the $u^{th}$ URLLC device \\\hline
			$W^{RB}$ & The total bandwidth of one resource block in the NB-IoT \\\hline
			$W$ & The bandwidth of one tone in one RB \\\hline
			$h^{s}_{m}$ &  The channel gain of the $m^{th}$ mMTC device over the $s^{th}$ sub-channel\\\hline
			$h^{s}_{u}$ &  The channel gain of the $u^{th}$ URLLC device over the $s^{th}$ sub-channel\\\hline
			$R^{th}_{m}$ &  The minimum transmission rate of the $m^{th}$ mMTC device \\\hline
			$R^{th}_{u}$ &  The minimum transmission rate of the $u^{th}$ URLLC device \\\hline
			$P^{max}_{m}$ &  The maximum power budget of the $m^{th}$ mMTC device \\\hline
			$P^{max}_{u}$ &  The maximum power budget of the $u^{th}$ URLLC device \\
			\hline
		\end{tabular}
	\end{table}

Denote $p_m^s$ as the transmission power of the $m^{th}$ mMTC over the $s^{th}$ sub-channel and $\alpha_m^{c,k}$ as the binary variable to allocate of the $m^{th}$ mMTC to the $k^{th}$ order of cluster $c$. In fact, $\alpha _m^{c,k} = 1$ if there is an allocation, and $\alpha _m^{c,k} = 0$ otherwise. Thus, the achievable data rate of the $m^{th}$ mMTC device, $R_m$, in terms of the aggregate rate over the allocated sub-carriers can be expressed as

\begin{equation}
	\begin{gathered}
		{R_m} = \sum\limits_{c \in \mathcal{C}} {\sum\limits_{k \in \mathcal{K}} {\alpha _m^{c,k}\sum\limits_{s \in \mathcal{S}} {{\gamma ^{s,c}}W} } }  \hfill \\
		{\log _2}\left( {1 + \frac{{{{\left| {h_m^s} \right|}^2}p_m^s}}{{{N_0}W + \sum\limits_{d \in \mathcal{M}\backslash m} {\sum\limits_{h = k + 1}^{{k_{\max }}} {\alpha _d^{c,h}{{\left| {h_d^s} \right|}^2}p_d^s} } }}} \right), \hfill \\ 
	\end{gathered}
\end{equation}
where $N_0$ is the noise power spectral density and $h_m^s$ denotes the channel gain between the $m^{th}$ mMTC device and the eNB on sub-channel $s$. Since the NOMA clustering procedure requires mMTC devices to have higher ranks as compared to URLLCs, the URLLC devices do not interfere mMTCs. Thus, the $m^{th}$ mMTC only experiences interference from the other mMTCs of the same cluster with higher ranks.

Note that each mMTC device requires a threshold for its data rate to be greater than the minimal data rate of $R_m^{th}$, i.e.,

\begin{equation} \label{Rm}
	{R_m} \geqslant {R_m^{th}},~~\forall m \in \mathcal{M}.
\end{equation}
The total transmission power of the $m^{th}$ mMTC device is limited to its maximum power budget $P_m^{max}$, i.e.,

\begin{equation} \label{Pm}
	\sum\limits_{s \in \mathcal{S}} {p_m^s}  \leqslant P_m^{\max },~~\forall m \in \mathcal{M}.
\end{equation}
Similarly, the achievable data rate of the $u^{th}$ URLLC device can be determined by the Shannon-Hartley theorem. Note that the ranks of URLLCs are always greater than those of mMTCs in each NOMA cluster. Thus, they receive interference from all the mMTC cluster members as well as those URLLC cluster members with higher ranks. Denote $\beta _u^{c,k}$ as the binary variable to assign the $u^{th}$ URLLC to the $k^{th}$ order of cluster $c$. In other words, $\beta _u^{c,k} = 1$ if there is an allocation, and $\beta _u^{c,k} = 0$ otherwise. Therefore, the achievable data rate of the $u^{th}$ URLLC device over the allocated sub-carriers is shown in Eq.~(\ref{longeq}), where $h_u^s$ is the channel gain between the $u^{th}$ URLLC device and the eNB on sub-channel $s$, and $p_u^s$ represents the transmission power of the $u^{th}$ URLLC over the $s^{th}$ sub-channel. Owing to performing critical tasks by URLLC devices, their power consumption is not of significant importance. Thus, the transmission powers of URLLC devices are set to their maximum limit, i.e.,
\begin{equation}\label{longeq}
		{R_u} = \sum\limits_{c \in \mathcal{C}} {\sum\limits_{k \in \mathcal{K}} {\beta _u^{c,k}\sum\limits_{s \in \mathcal{S}} {{\gamma ^{s,c}}W} } } {\log _2}\left( {1 + \frac{{{{\left| {h_u^s} \right|}^2}p_u^s}}{{{N_0}W + \sum\limits_{d \in \mathcal{U}\backslash u} {\sum\limits_{h = k + 1}^{{k_{\max }}} {\beta _d^{c,h}{{\left| {h_d^s} \right|}^2}p_d^s + \sum\limits_{m \in \mathcal{M}} {\sum\limits_{h = k + 1}^{{k_{\max }}} {\alpha _m^{c,h}{{\left| {h_m^s} \right|}^2}p_m^s} } } } }}} \right),
	\end{equation}
\begin{equation} \label{Pu}
	\sum\limits_{s \in \mathcal{S}} {p_u^s}  = P_u^{\max },~~\forall u \in \mathcal{U}.
\end{equation}
Meanwhile, the data rate of the $u^{th}$ URLLC device should be greater than a given minimal rate $R_u^{th}$,

\begin{equation} \label{Ru}
	{R_u} \geqslant {R_u^{th}},~~\forall u \in \mathcal{U}.
\end{equation}

\section{The optimization framework}

In this section, the optimization problem of NOMA clustering for NB-IoT is formulated as a sum rate maximization of URLLC and mMTC devices. Apart from the QoS constraints in Eq.~(\ref{Rm}), Eq.~(\ref{Pm}), Eq.~(\ref{Pu}), and Eq.~(\ref{Ru}), we should enforce extra constraints for the NOMA clustering process. In particular, each URLLC and mMTC device should be assigned to only one cluster with one specific rank, i.e.,

\begin{equation}
	\sum\limits_{c \in \mathcal{C}} {\sum\limits_{k \in \mathcal{K}} {\alpha _m^{c,k}}  = 1,~~~\forall m \in \mathcal{M}},  
\end{equation}

\begin{equation}
	\sum\limits_{c \in \mathcal{C}} {\sum\limits_{k \in \mathcal{K}} {\beta _u^{c,k}}  = 1,~~~\forall u \in \mathcal{U}}. 
\end{equation}
Moreover, each rank of one cluster should be assigned either to one URLLC or one mMTC, i.e.,

\begin{equation}
	\sum\limits_{m \in \mathcal{M}} {\alpha _m^{c,k} + } \sum\limits_{u \in \mathcal{U}} {\beta _u^{c,k}}  = 1,~~\forall c \in \mathcal{C},~~\forall k \in \mathcal{K}. 
\end{equation}
Since NOMA is to share spectral resources between multiple users, the NOMA clustering enforces existence of more than one user in each cluster, i.e.,

\begin{equation}
	\sum\limits_{m \in \mathcal{M}} {\sum\limits_{k \in \mathcal{K}} {\alpha _m^{c,k} + \sum\limits_{u \in \mathcal{U}} {\sum\limits_{k \in \mathcal{K}} {\beta _u^{c,k} \geqslant 2} } ,~~~\forall c \in \mathcal{C}} }. 
\end{equation}
Note that the order of users in each cluster $c \in \mathcal{C}$ can significantly affect the network throughput. The URLLC devices are prioritized to have the lowest ranks of clusters (i.e., $k = 1,2, ... $) due to their higher data rate and transmission power requirements. In other words, the high power of URLLCs do not affect the low power mMTC devices during the SIC process, if they are assigned to the lowest ranks of clusters. Therefore, for the $k^{th}$ ($2 \leqslant k \leqslant {k_{\max }}$) rank of each cluster, the mMTC devices should always have higher ranks than the URLLC devices, i.e.,

\begin{equation}
	\beta _u^{c,k} \geqslant \alpha _m^{c,k - 1},~~\forall m \in \mathcal{M},~~\forall u \in \mathcal{U},~~\forall c \in \mathcal{C},
\end{equation}
and we ensure the ranks' assignment priority in each cluster, by starting rank assignments from the lowest rank of each cluster ($k=1$), i.e., 

\begin{equation}
	\alpha _m^{c,k} \leqslant \alpha _m^{c,k - 1},~\forall m \in \mathcal{M},~\forall c \in \mathcal{C},~2 \leqslant k \leqslant {k_{\max }},
\end{equation}

\begin{equation}
	\beta _u^{c,k} \leqslant \beta _u^{c,k - 1},~\forall u \in \mathcal{U},~\forall c \in \mathcal{C},~2 \leqslant k \leqslant {k_{\max }}.
\end{equation}
Finally, the NOMA clustering optimization problem for NB-IoT as a sum rate maximization of URLLC and mMTC devices can be given as

\begin{equation} \label{Opt}
	\begin{gathered}
		\textbf{P1:}~~\mathop {\max }\limits_{p_m^s,p_u^s,\alpha _m^{c,k},\beta _u^{c,k},{\gamma ^{s,c}}} \sum\limits_{m \in \mathcal{M}} {{R_m} + } \sum\limits_{u \in \mathcal{U}} {{R_u}}  \hfill \\
		s.t. \hfill \\
		C1:{R_m} \geqslant R_m^{th},~~\forall m \in \mathcal{M}, \hfill \\
		C2:\sum\limits_{s \in \mathcal{S}} {p_m^s}  \leqslant P_m^{\max },~~\forall m \in \mathcal{M}, \hfill \\
		C3:{R_u} \geqslant R_u^{th},~~\forall u \in \mathcal{U}, \hfill \\
		C4:\sum\limits_{s \in \mathcal{S}} {p_u^s}  = P_u^{\max },~~\forall u \in \mathcal{U}, \hfill \\
		C5:\beta _u^{c,k} \geqslant \alpha _m^{c,k - 1},~~\forall m \in \mathcal{M},~~\forall u \in \mathcal{U},~~\forall c \in \mathcal{C}, \hfill \\
		~~~~~~2 \leqslant k \leqslant {k_{\max }}, \hfill \\
		C6:\alpha _m^{c,k} \leqslant \alpha _m^{c,k - 1},~\forall m \in \mathcal{M},~\forall c \in \mathcal{C},~2 \leqslant k \leqslant {k_{\max }} \hfill \\
		C7:\beta _u^{c,k} \leqslant \beta _u^{c,k - 1},~\forall u \in \mathcal{U},~\forall c \in \mathcal{C},~2 \leqslant k \leqslant {k_{\max }} \hfill \\ 
		C8:\sum\limits_{c \in \mathcal{C}} {\sum\limits_{k \in \mathcal{K}} {\alpha _m^{c,k}}  = 1,~~\forall m \in \mathcal{M}},  \hfill \\
		C9:\sum\limits_{c \in \mathcal{C}} {\sum\limits_{k \in \mathcal{K}} {\beta _u^{c,k}}  = 1,~~\forall u \in \mathcal{U}},  \hfill \\
		C10:\sum\limits_{m \in \mathcal{M}} {\alpha _m^{c,k} + } \sum\limits_{u \in \mathcal{U}} {\beta _u^{c,k}}  = 1,~~\forall c \in \mathcal{C},~~\forall k \in \mathcal{K}, \hfill \\
		C11:\sum\limits_{m \in \mathcal{M}} {\sum\limits_{k \in \mathcal{K}} {\alpha _m^{c,k} + \sum\limits_{u \in \mathcal{U}} {\sum\limits_{k \in \mathcal{K}} {\beta _u^{c,k} \geqslant 2} } ,~~\forall c \in \mathcal{C}} },  \hfill \\
		C12:\sum\limits_{c \in \mathcal{C}} {{\gamma ^{s,c}} = 1,~~\forall s \in \mathcal{S}},  \hfill \\
		C13: \sum\limits_{s \in \mathcal{S}} {\sum\limits_{c \in \mathcal{C}} {{\gamma ^{s,c}}{\mathcal{W}^{s,c}}} }  \leqslant {W^{RB}},~~\forall c \in \mathcal{C},~~\forall s \in \mathcal{S} \hfill \\
		C14:p_m^s \geqslant 0,~~\forall m \in \mathcal{M},~~\forall s \in \mathcal{S}, \hfill \\
		C15:p_u^s \geqslant 0,~~\forall u \in \mathcal{U},~~\forall s \in \mathcal{S}, \hfill \\
		C16:{\gamma ^{s,c}} \in \left\{ {0,1} \right\},~~\forall c \in \mathcal{C},~~\forall s \in \mathcal{S}, \hfill \\
		C17:\alpha _m^{c,k} \in \left\{ {0,1} \right\},~~\forall m \in \mathcal{M},~~\forall c \in \mathcal{C},~~\forall k \in \mathcal{K}, \hfill \\
		C18:\beta _u^{c,k} \in \left\{ {0,1} \right\},~~\forall u \in \mathcal{U},~~\forall c \in \mathcal{C},~~\forall k \in \mathcal{K}, \hfill \\ 
	\end{gathered}
\end{equation}
where C1 imposes the data rates of mMTC devices to be greater than the minimum data rate requirement. C2 limits the total transmission power of the $m^{th}$ mMTC to the maximum power budget, $P_m^{max}$. C3 implies that the minimum data rate constraint for each URLLC device must be satisfied. C4 is the power budget constraint for each URLLC device. C5 is to ensure that the ranks of mMTC devices are higher than those of URLLCs for each NOMA cluster. C6 and C7 imply that mMTC and URLLC devices can be assigned to the $k^{th}$ rank of the $c^{th}$ cluster if all the lower ranks are already allocated to other users. C8 and C9 are designed to guarantee that each device (mMTC and URLLC) is allocated to only one cluster and one specific order within the cluster. C10 specifies that each rank of a cluster cannot be allocated to both mMTC and URLLC devices. C11 is to guarantee each NOMA cluster to have more than one member. C12 implies that each sub-carrier cannot be allocated to more than one cluster. 
C13 ensures that the total bandwidth allocated to all NOMA clusters is not more than the bandwidth of one RB (bandwidth of one RB in NB-IoT is 180 kHz). C14  and C15 are to limit the transmission powers of mMTCs and URLLCs to positive values. C16, C17 and C18 ensure that the variables $\gamma^{s,c}$, $\alpha_m^{c,k}$, and $\beta_u^{c,k}$ are restricted to binary values, respectively.

\begin{lemma} \label{[NP-Hard]}
	The general optimization problem of NOMA clustering problem for NB-IoT in Eq.~(\ref{Opt}) is an NP-hard problem.
\end{lemma}
\begin{IEEEproof}
	Without loss of generality, we assume that URLLC and mMTC users are assigned to different clusters with various ranks in the clusters. Therefore, the values of $\alpha_m^{c,k}$, and $\beta_u^{c,k}$ are determined and the corresponding constraints in \textbf{P1} are relaxed. Given that URLLC and mMTC users transmit their data with predetermined transmission powers of $p_u^s$ and $p_m^s$, the constraints related to these two variables are relaxed and the NOMA clustering optimization problem for NB-IoT as a sum rate maximization of URLLC and mMTC devices is reduced to the following:
	
	\begin{equation} \label{Opt2}
		\begin{gathered}
			\textbf{P2:}~~\mathop {\max }\limits_{{\gamma ^{s,c}}} \sum\limits_{s \in \mathcal{S}} {\sum\limits_{c \in \mathcal{C}} {{\gamma ^{s,c}}} \left( {\sum\limits_{m \in \mathcal{M}} {R_m^{s,c} + } \sum\limits_{u \in \mathcal{U}} {R_u^{s,c}} } \right)}  \hfill \\
			s.t. \hfill \\
			C1: \sum\limits_{s \in \mathcal{S}} {\sum\limits_{c \in \mathcal{C}} {{\gamma ^{s,c}}{\mathcal{W}^{s,c}}} }  \leqslant {W^{RB}},~~\forall c \in \mathcal{C},~~\forall s \in \mathcal{S} \hfill \\
			C2: \sum\limits_{c \in \mathcal{C}} {{\gamma ^{s,c}}}  = 1,~~\forall c \in \mathcal{C},~~\forall s \in \mathcal{S} \hfill \\
			C3: {\gamma ^{s,c}} \in \left\{ {0,1} \right\},~~\forall c \in \mathcal{C},~~\forall s \in \mathcal{S} \hfill \\
		\end{gathered}
	\end{equation}
	
	Hence, the reduced optimization problem, \textbf{P2}, is similar to a Multiple Choice Knapsack Problem (MCKP). In fact, the problem would be the problem of packing $\left| \mathcal{S} \right|$ items (sub-channels) into $\left| \mathcal{K} \right|$ knapsacks (clusters). Each item (sub-channel), \textit{s},  has a weight if allocated to the $c^{th}$ knapsack (cluster). Moreover, each sub-channel has a profit which is $({\sum\limits_{m \in \mathcal{M}} {R_m^{s,c} + } \sum\limits_{u \in \mathcal{U}} {R_u^{s,c}} })$ and the problem is to choose items such that the profit sum is maximized without exceeding the capacity, ${W^{RB}}$. Therefore, \textbf{P2} is NP-hard because it is categorized as a MCKP which is a generalization of the ordinary knapsack problem. Thus, as \textbf{P2} is a special case of \textbf{P1}, the general optimization problem in Eq.~(\ref{Opt}) is an NP-hard problem.
\end{IEEEproof}

The formulated optimization problem is a non convex mixed integer nonlinear programming (MINLP) problem which is combinatorial, and exploiting exhaustive search presents exponential time complexity. Therefore, we solve the optimization problem by proposing a heuristic algorithm.

\section{Proposed Algorithm}

\begin{algorithm}[H] \label{algorithm}
	\textbf{Initialization}\\
	Input: $C$, $R_m^{th}$, $R_u^{th}$, $P_m^{max}$, $P_u^{max}$, $h_m^{s}$, and $h_u^{s}$,\\ $\forall m \in \mathcal{M}$, $\forall u \in \mathcal{U}$, $\forall s \in \mathcal{S}$\\
	\textbf{URLLC Clustering}\\
	Sorting URLLC devices $\forall u \in \mathcal{U}$: ${{\tilde h}_1} \geqslant {{\tilde h}_2} \geqslant ... \geqslant {{\tilde h}_U}$ \\
	\textit{\textbf{for all}} $u \in \mathcal{U}$ \textit{\textbf{do}}\\
	\textit{\textbf{if}}  $U \le C$ \textit{\textbf{do}}\\
	Assign URLLC devices $\{ 1,2,...,U\}$ to the lowest rank ($k=1$) of $\{ 1,2,...,C\}$ clusters\\
	\textit{\textbf{else}}\\
	Assign URLLC $\{ 1,2,...,C\}$ to the lowest rank of all $C$ clusters, and $\{ C + 1,C + 2,...,U\}$ to the higher ranks\\
	\textit{\textbf{end if}}\\	 
	\textit{\textbf{end for}}\\
	\textbf{mMTC Clustering}\\
	Sorting mMTC devices $\forall m \in \mathcal{M}$: ${{\tilde h}_1} \geqslant {{\tilde h}_2} \geqslant ... \geqslant {{\tilde h}_M}$ \\
	\textit{\textbf{for all}} $k \in \mathcal{K}$ \textit{\textbf{do}}\\
	\textit{\textbf{if}}  $U < C$ \textit{\textbf{do}}\\
	Assign mMTC $\{ 1,...,(C-U)\}$ to the lowest rank ($k=1$) of $\{ (U+1),...,C\}$ clusters.\\
	\textit{\textbf{else}}\\
	Assign mMTC $\{ 1,...,(C-U)\}$ to the next available rank of $\{(U+1),...,C\}$ clusters.\\
	\textit{\textbf{end if}}\\	
	\textit{\textbf{end for}}\\
	\textbf{Resource Allocation for NOMA Clusters}\\
	Set ${R_u} = 0$, ${R_m} = 0$, $p_m^{s}=P_m^{max}$ and $p_u^{s}= P_u^{max}$, $\forall m \in \mathcal{M}$, $\forall u \in \mathcal{U}$, $\forall s \in \mathcal{S}$, $ {\hat S} \leftarrow \emptyset ,~~S_a^c \leftarrow \emptyset ,~~{C_{ns}} \leftarrow \mathcal{C}$\\
	{\textit{\textbf{While}} ${{\cal S}} \ne \emptyset $ and ${R_u} < R_u^{th}$ and ${R_m} < R_m^{th}$}\\{
		Select the best cluster $c^*$, $\forall c \in \mathcal{C}$, for each sub-carrier $s \in \mathcal{S}$:\\
		${c^*} = \mathop {\arg \max }\limits_{c \in {C_{ns}}} \left( {\sum\nolimits_{u \in \mathcal{U}} {{R_u} + \sum\nolimits_{m \in \mathcal{M}} {{R_m}} } } \right)$;\\
		Allocate the sub-carrier $s$ to the cluster $c^*$:\\
		Set ${\gamma ^{s,{c^*}}} = 1$, and update $S_a^{c^*} \leftarrow S_a^{c^*} \cup \{ s\}$, $\hat S \leftarrow \hat S \cup \{ s\}$\\
		Update the rates: ${R_u} = {R_u} + {R_{u,s}}$, ${R_m} = {R_m} + {R_{m,s}}$\\
		Update the powers: URLLC and mMTC of $c^*$ individually perform SUWF over all allocated sub-carriers: \\
		$p_m^s = \frac{{p_m^s}}{{\left| {S_a^{c^*}} \right| + 1}}$, $p_u^s = \frac{{p_u^s}}{{\left| {S_a^{c^*}} \right| + 1}}$, $\forall s \in \mathcal{S}$ \\
		\textit{\textbf{if}} ${R_u} \geqslant R_u^{th}$ and ${R_m} \geqslant R_m^{th}$, $\forall m,u$ from cluster $c^*$ \textit{\textbf{do}}\\
		${C_{ns}} \leftarrow {C_{ns}}\backslash \{{c^*}\}$\\
		\textit{\textbf{end if}}\\
		$\mathcal{S} \leftarrow \mathcal{S}\backslash \hat S$\\
		\textit{\textbf{if}} ${R_u} \geqslant R_u^{th}$ and ${R_m} \geqslant R_m^{th}$, $\forall m \in \mathcal{M}$, $\forall u \in \mathcal{U}$ \textit{\textbf{do}}\\
		\textit{\textbf{for all}} $s \in \mathcal{S}$ \textit{\textbf{do}}\\
		${c^*} = \mathop {\arg \max }\limits_{c \in {C}} \left( {\sum\nolimits_{u \in \mathcal{U}} {{R_u} + \sum\nolimits_{m \in \mathcal{M}} {{R_m}} } } \right)$\\
		Set ${\gamma ^{s,{c^*}}} = 1$, $S_a^{c^*} \leftarrow S_a^{c^*} \cup \{ s\}$\\
		\textit{\textbf{end for}}\\
		Update $p_m^s = \frac{{p_m^s}}{{\left| {S_a^{c^*}} \right| + 1}}$, $p_u^s = \frac{{p_u^s}}{{\left| {S_a^{c^*}} \right| + 1}}$\\
		\textit{\textbf{end if}}\\
		\textit{\textbf{End while}}
	}
	\caption{NOMA Clustering and Resource Allocation for Machine Type Communication}
\end{algorithm}
In this section, we propose an efficient heuristic algorithm to find sub-optimal solutions of the non convex MINLP problem in Eq.~(\ref{Opt}). The proposed algorithm optimizes the NOMA clustering of mMTC and URLLC devices and allocates spectral resources to the NOMA clusters. The pseudo code for solving the optimization problem is summarized in Algorithm 1. The first phase of the algorithm is the URLLC clustering, where the URLLC devices are sorted based on their average channel gains, ${{\tilde h}_u} = {\raise0.5ex\hbox{$\scriptstyle {\sum\limits_{s \in \mathcal{S}} {h_u^s} }$}
	\kern-0.1em/\kern-0.15em
	\lower0.25ex\hbox{$\scriptstyle S$}}$.
As discussed in Subsection~\ref{NomaClustering}, the URLLC devices have higher data rate and transmission power requirements. Therefore, to mitigate the adverse impacts of interference caused by the URLLCs' high transmission powers, the ranks of URLLC devices in each cluster should be lower than the mMTC ones. In the URLLC clustering process, URLLC devices with higher ${\tilde h}_u$ are assigned to the lowest ranks of NOMA clusters, i.e., $k=1$. If the number of URLLC devices, $U$, is greater than the number of NOMA clusters, $C$, the remaining devices are assigned to the next ranks of clusters. Similar to the URLLC clustering approach, the mMTC clustering procedure is based on the average channel gain of mMTC devices, ${{\tilde h}_m} = {\raise0.5ex\hbox{$\scriptstyle {\sum\limits_{s \in \mathcal{S}} {h_m^s} }$}
	\kern-0.1em/\kern-0.15em
	\lower0.25ex\hbox{$\scriptstyle S$}}$.
The mMTC devices with higher ${\tilde h}_m$ are allocated to the next available rank of clusters. Then, the remaining mMTC devices are allocated to the higher ranks of NOMA clusters. By this NOMA clustering approach, constraints 5-11 in Eq.~(\ref{Opt}) are taken into consideration. After the NOMA clustering process, the resource allocation for URLLC and mMTC devices are detailed in Algorithm 1. The initial values for the transmission rates and powers of URLLC and mMTC devices are ${R_u} = 0$, $p_u^{s}= P_u^{max}$, and ${R_m} = 0$, $p_m^{s}=P_m^{max}$, respectively.  The resource allocation phase continues until all the sub-channels are allocated to NOMA clusters and the data rate requirements of mMTC and URLLC devices are satisfied.
Denote $S_a^c \leftarrow \emptyset$ as the set of allocated sub-channels to the $c^{th}$ cluster, and ${C_{ns}} \leftarrow \mathcal{C}$ as the set of clusters of devices with unsatisfied rates. For each sub-carrier, the best cluster ($c^*$) is the one that maximizes the total throughput, i.e., ${c^*} = \mathop {\arg \max }\limits_{c \in {C_{ns}}} \left( {\sum\nolimits_{u \in \mathcal{U}} {{R_u} + \sum\nolimits_{m \in \mathcal{M}} {{R_m}} } } \right)$. Then, the data rates of the mMTC and URLLC devices and their transmission powers are updated accordingly. Note that each MTC device performs Single User Water Filling (SUWF)\cite{TafazolliNOMA} technique over all allocated sub-channels. During the resource allocation process, clusters with satisfied data rates are excluded from the set of $C_{ns}$. The algorithm iteratively allocates the sub-channels one by one until all the mMTC and URLLC devices' rate requirements are met.

\subsection{Power Allocation}

Given the URLLC and mMTC user allocation to NOMA clusters and spectrum allocation to the clusters, the binary variables of $\alpha _m^{c,k}$, $\beta _u^{c,k}$ and ${{\gamma ^{s,c}}}$ in \textbf{P1} take on 0 or 1. Therefore, all integer constraints are removed and the new optimization problem, which tries to find optimal values of URLLC and mMTC transmission powers, can be expressed as

\begin{equation} \label{Opt3}
	\begin{gathered}
		\textbf{P3:}~~\mathop {\max }\limits_{p_m^s,p_u^s} \sum\limits_{m \in \mathcal{M}} {{R_m} + } \sum\limits_{u \in \mathcal{U}} {{R_u}}  \hfill \\
		s.t. \hfill \\
		C1,~C2,~C3,~C4,~C14,~\text{and}~C15~\text{in}~\textbf{P1} \hfill\\
	\end{gathered}
\end{equation}

The reduced optimization problem, given its original formulation in \textbf{P3}, is apparently non-convex due to the interference users introduce to each other. To address this, we first define a new set of both URLLC and mMTC users, $\mathcal{J} = \left\{ {1,2,...,U,U + 1,...,U + M} \right\}$ for one cluster (the result is also valid for more clusters). Let ${\lambda _j} \triangleq \tfrac{{{{\left| {{h_j}} \right|}^2}}}{{{N_0}W}}$, where $h_j$ is the channel coefficient from the $j^{th}$ user to the eNB. Without loss of generality, we order users by their normalized channel gains as ${\lambda _1} \leqslant {\lambda _2} \leqslant ... \leqslant {\lambda _{U + M}}$. Note that users exploit SIC at their receivers such that ${P_1}\geqslant{P_2}\geqslant...\geqslant{P_U}\geqslant{P_{U + 1}}\geqslant...\geqslant{P_{U + M}}$, where ${P_j} \triangleq \sum\limits_{s \in \mathcal{S}} {p_j^s}$. Therefore, \textbf{P3} can be rewritten as

\begin{equation} \label{Opt4}
	\begin{gathered}
		\textbf{P4:}~~\mathop {\max }\limits_{{P_j}} \sum\limits_{j \in \mathcal{J}} {{R_j}}  \hfill \\
		s.t. \hfill \\
		C1:{R_j} \geqslant R_j^{th},~~\forall j \in \mathcal{J}, \hfill \\
		C2:\sum\limits_{j \in \mathcal{J}} {{P_j}} \leqslant {P^{\max }},~~\forall j \in \mathcal{J}, \hfill \\
		C3:{P_1}\geqslant{P_2}\geqslant...\geqslant{P_U}\geqslant{P_{U + 1}}\geqslant...\geqslant{P_{U + M}}, \hfill \\
	\end{gathered}
\end{equation}
where ${R_j} \triangleq W^{RB} {\log _2}(1 + \tfrac{{{\lambda _j}{P_j}}}{{1 + {\lambda _j}\sum\nolimits_{l = j + 1}^{U + M} {{P_l}} }})$. To make \textbf{P4} convex, we use the variable transformations of ${Z_j} = \sum\nolimits_{l = j}^{U + M} {{P_l}} ,~\forall j \in \mathcal{J}$, or ${P_j} = {Z_j} - {Z_{j + 1}},~\forall j \in \left\{ {1,2,...,U + M - 1} \right\}$ and ${P_{U + M}} = {Z_{U + M}}$. Therefore, we can rewrite $R_j, ~\forall j \in \left\{ {1,2,...,U + M - 1} \right\}$ as

\begin{equation}
	\begin{gathered}
		{R_j} = {\log _2}\left( {1 + \tfrac{{{\lambda _j}{P_j}}}{{1 + {\lambda _j}\sum\nolimits_{l = j + 1}^{U + M} {{P_l}} }}} \right) = {\log _2}\left( {\tfrac{{1 + {\lambda _j}\sum\nolimits_{l = j}^{U + M} {{P_l}} }}{{1 + {\lambda _j}\sum\nolimits_{l = j + 1}^{U + M} {{P_l}} }}} \right) \\
		=  {\log _2}\left( {\tfrac{{1 + {\lambda _j}{Z_j}}}{{1 + {\lambda _j}{Z_{j + 1}}}}} \right) = {\log _2}\left( {1 + {\lambda _j}{Z_j}} \right) - {\log _2}\left( {1 + {\lambda _j}{Z_{j + 1}}} \right),
	\end{gathered}
\end{equation}
while for $j=U+M$, ${R_{U + M}} = {\log _2}\left( {1 + {\lambda _{U + M}}{Z_{U + M}}} \right)$. Thus, the objective function in \textbf{P3} ($\sum\nolimits_{j = 1}^{U + M} {{R_j}} $) can be written as 

\begin{equation}
	\begin{gathered}
		\sum\nolimits_{j = 1}^{U + M - 1} {{W^{RB}}\left[ {{{\log }_2}\left( {1 + {\lambda _j}{Z_j}} \right) - {{\log }_2}\left( {1 + {\lambda _j}{Z_{j + 1}}} \right)} \right]}\\  + {W^{RB}}{\log _2}\left( {1 + {\lambda _{U + M}}{Z_{U + M}}} \right) = \sum\nolimits_{j = 1}^{U + M} {{\Phi _j}({Z_j})},
	\end{gathered}
\end{equation}
where ${\Phi _1}({Z_1}) \triangleq {W^{RB}}{\log _2}\left( {1 + {\lambda _1}{Z_1}} \right)$, and for all $j \in \left\{ {2,3,...,U + M} \right\}$,

\begin{equation}
	{\Phi _j}({Z_j}) \triangleq {W^{RB}}\left[ {{{\log }_2}\left( {1 + {\lambda _j}{Z_j}} \right) - {{\log }_2}\left( {1 + {\lambda _{j - 1}}{Z_j}} \right)} \right].
\end{equation}
The rate constraint, C1 in \textbf{P4}, can be linearized by using ${Z_{j + 1}} \leqslant {\delta _j}{Z_j} - {\rho _j}$ for all $j \in \left\{ {1,2,...,U + M-1} \right\}$, and ${Z_{U + M}} \geqslant {\theta _{U + M}}$, where ${\delta _j} \triangleq {2^{ - R_j^{th}}}$,
${\rho _j} \triangleq \tfrac{{(1 - {\delta _j})}}{{{\lambda _j}}}$, and ${\theta _j} \triangleq \tfrac{{({2^{R_j^{th}}} - 1)}}{{{\lambda _j}}}$. The transmission power in C2 of \textbf{P4} can be equivalent to ${Z_1} = \sum\nolimits_{j = 1}^{U + M} {{P_j} = {P^{\max }}}$. The power order constraint, C3 in \textbf{P4}, ${P_1}\geqslant{P_2}\geqslant...\geqslant{P_{U + M}}\geqslant 0$ is equivalent to ${Z_1} - {Z_2} \geqslant {Z_2} - {Z_3} \geqslant ... \geqslant {Z_{U + M}} \geqslant 0$. Therefore, the power allocation  problem in \textbf{P4} can be transformed to the following optimization problem 

\begin{equation} \label{Opt5}
	\begin{gathered}
		\textbf{P5:}~~\mathop {\max }\limits_{{\textbf{Z}}} {\sum\limits_{j \in \mathcal{J}} {{\Phi _j}({Z_j})}}  \hfill \\
		s.t. \hfill \\
		C1:{Z_{j + 1}} \leqslant {\delta _j}{Z_j} - {\rho _j}, \hfill \\
		C2:{Z_1} = {P^{\max }}, \hfill \\
		C3:{Z_1} - {Z_2} \geqslant {Z_2} - {Z_3} \geqslant ... \geqslant {Z_{U + M}} \geqslant {\theta _j}, \hfill \\
	\end{gathered}
\end{equation}
where $\textbf{Z} \triangleq \left( {{Z_j}} \right)_{j = 1}^{U + M}$. Note that the transformation between $P$ and $Z$ is linear, and therefore the convexity of \textbf{P3} is equivalent to the convexity of \textbf{P5}.

\begin{theorem}
	Given ${\lambda _1} \leqslant {\lambda _2} \leqslant ... \leqslant {\lambda _{U + M}}$, the power allocation problem in \textbf{P3} (or equivalently \textbf{P5}) is a convex optimization problem, for all $j \in \left\{ {2,3,...,U + M} \right\}$.
\end{theorem}

\begin{IEEEproof}
	We start to prove the theorem by investigating the objective function of \textbf{P5} (${{\Phi _j}({Z_j})}$) due to the fact that all constraints are linear.  The derivative of the objective function for all $j \in \left\{ {2,3,...,U + M} \right\}$ is given by 
	
	\begin{equation}
		\frac{{{\Phi _j}({Z_j})}}{{d{Z_j}}} = \frac{{{\lambda _j}}}{{1 + {\lambda _j}{Z_j}}} - \frac{{{\lambda _{j - 1}}}}{{1 + {\lambda _{j - 1}}{Z_j}}}.
	\end{equation}
	The second derivative of ${{\Phi _j}({Z_j})}$ is given by
	
	\begin{equation}
		\begin{gathered}
			\Phi _j^{''}({Z_j}) = \frac{{ - {{({\lambda _j})}^2}}}{{{{(1 + {\lambda _j}{Z_j})}^2}}} - \frac{{ - {{({\lambda _{j - 1}})}^2}}}{{{{(1 + {\lambda _{j - 1}}{Z_j})}^2}}} \hfill \\
			= \frac{{\lambda _{j - 1}^2 - \lambda _j^2 + 2{\lambda _j}{Z_j}\lambda _{j - 1}^2 - 2{\lambda _{j - 1}}{Z_j}\lambda _j^2}}{{{{(1 + {\lambda _j}{Z_j})}^2}{{(1 + {\lambda _{j - 1}}{Z_j})}^2}}} \hfill \\ 
		\end{gathered} 
	\end{equation}
	Given ${\lambda _1} \leqslant {\lambda _2} \leqslant ... \leqslant {\lambda _{U + M}}$, the numerator of the second derivative is negative and the denominator is always positive. Therefore, the second derivative is negative and the objective function is concave.
\end{IEEEproof}

\section{Simulation Results}\label{sec:simultion}

In this section, we evaluate the system performance of the proposed NOMA based NB-IoT scheme with sub-carrier and power allocation, and the NOMA clustering via Monte Carlo simulation. We consider one cell with $0.5$ km radius where the locations of the mMTC and URLLC devices are randomly generated and uniformly distributed within the cell. We consider one PRB with 48 sub-carrier spacing of 3.75 kHz for the MTC uplink transmissions in one time slot. We model the channel gains of the mMTC devices as $h_{m}^s = {\mathcal{Y}}d_{m,s}^{-\beta }$ (similarly $h_u^s$ for URLLCs), where $\mathcal{Y}$ is a random value generated based on the Rayleigh distribution, $d_{m,s}^{-\beta}$ represents the distance between the transmitter and receiver, and $\beta$ is the path-loss exponent. We set $\beta=3$ and $d$ is varied between 0.1~m to 500~m. We also consider Additive White Gaussian Noise (AWGN) with power spectral density of -173 dBm/Hz. The maximum transmission power budgets of all URLLC and mMTC devices, $P_u^{max}$ and $P_m^{max}$ ($ \forall u \in \mathcal{U}, \forall m \in \mathcal{M}$), are set to 23 dbm. The data rate thresholds of the mMTC devices follow uniform distribution, i.e., $R_m^{th}=$Uniform $(0.1,2)$~kbps. The bandwidth of each sub-carrier in one PRB with 48 sub-carriers is set to $w=3.75$kHz. The Orthogonal Frequency Division Multiple Access (OFDMA) scheme as an OMA scheme and the fast OFDM \cite{DoubleConnectedNBIoT} approach are used for benchmark comparison.

\begin{figure} [t]
	\centering
	\includegraphics[width=12cm,height=8cm]{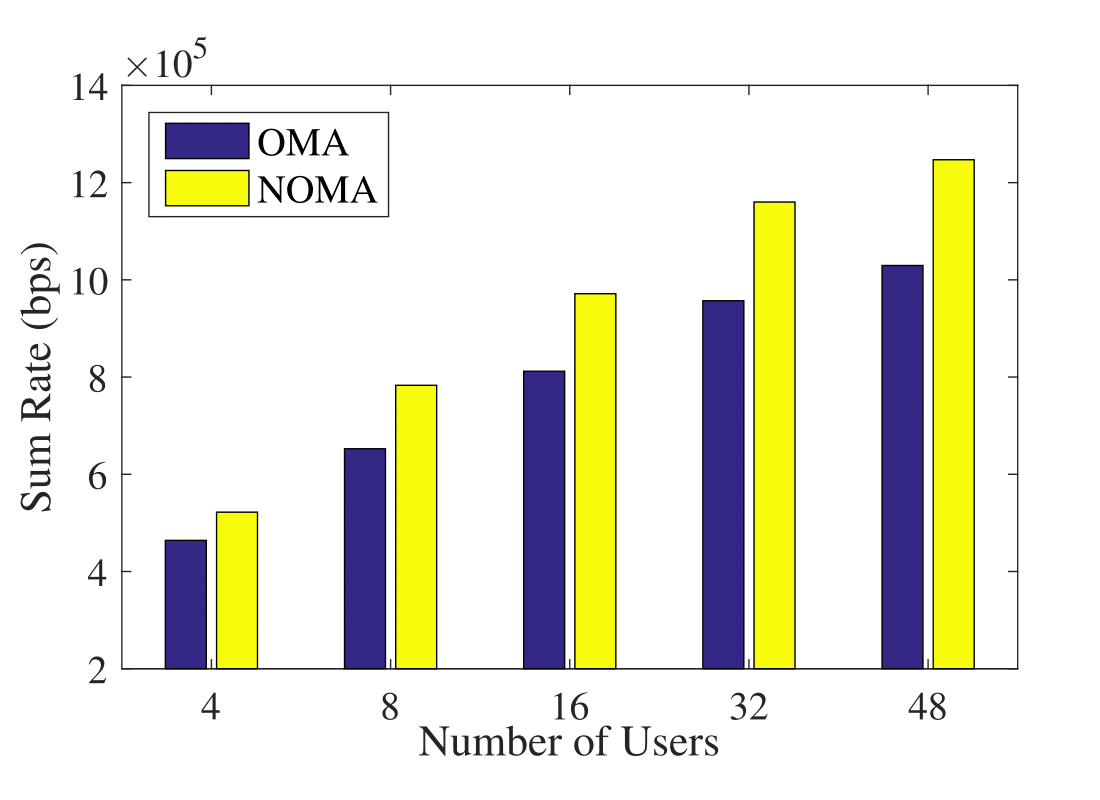}
	\caption{The total throughput of a NOMA based NB-IoT system with respect to the number of users (mMTC and URLLC devices).}
	\label{fig:SumRate}
\end{figure}

Fig.~{\ref{fig:SumRate}} compares the sum rate of the NOMA and the OMA schemes for an NB-IoT system with respect to the total number of mMTC and URLLC devices. As we can see in this figure, the performance gain in the total throughput for the proposed NOMA based NB-IoT scheme over the OMA scenario is approximately $28\%$ for a sufficiently large number of users. Owing to the multi-user diversity gain, the sum rate increases according to the number of users. Note that the ratio of the mMTC devices to the URLLC ones is set to 3, and the data rate thresholds of the URLLC devices are uniformly distributed between 0.1 kbps and 20 kbps.

\begin{figure} [t]
	\centering
	\includegraphics[width=12cm,height=8cm]{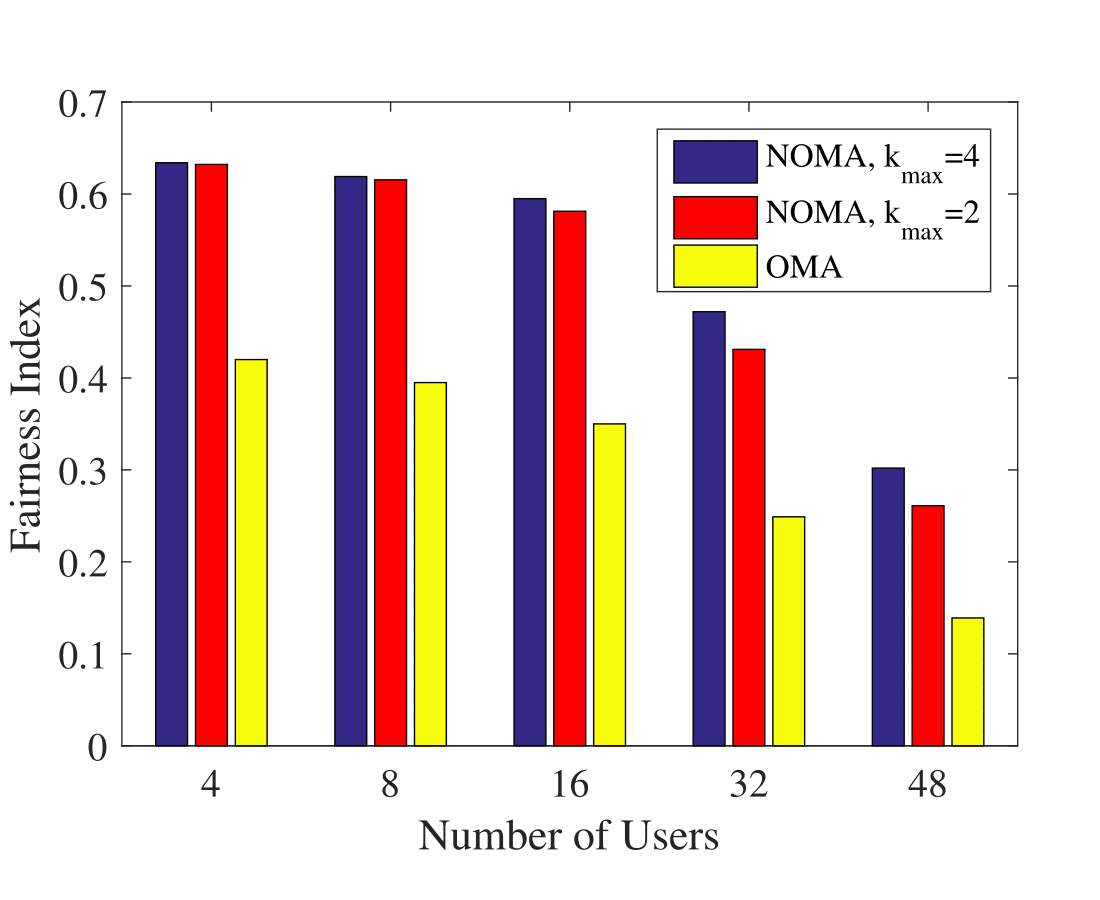}
	\caption{The fairness comparison between OMA and NOMA schemes.}
	\label{fig:Fairness}
\end{figure}

To compare the fairness of the proposed NOMA scheme and the OMA scenario, the Jain's fairness index~\cite{TafazolliNOMA} is adopted for data rates of mMTC and URLLC devices, i.e., Fairness Index$= \tfrac{{{{\left( {\sum\nolimits_{u = 1}^U {{R_u}}  + \sum\nolimits_{m = 1}^M {{R_m}} } \right)}^2}}}{{\left( {U + M} \right)\left( {\sum\nolimits_{u = 1}^U {R_u^2}  + \sum\nolimits_{m = 1}^M {R_m^2} } \right)}}$. In fact, Jain's fairness index is bounded between 0 and 1, and the maximum value is obtained if all the devices achieve exactly the same throughput. Fig.~\ref{fig:Fairness} shows the Jain's fairness index for both NOMA and OMA schemes. As the figure shows, the NOMA scheme for both $k_{max}=2$ and $k_{max}=4$ scenarios are fairer as compared to the OMA scheme because the OMA scheme does not allocate one sub-channel to more than one user, thus depriving some users from spectral resources.

\begin{figure} [t]
	\centering
	\includegraphics[width=12cm,height=8cm]{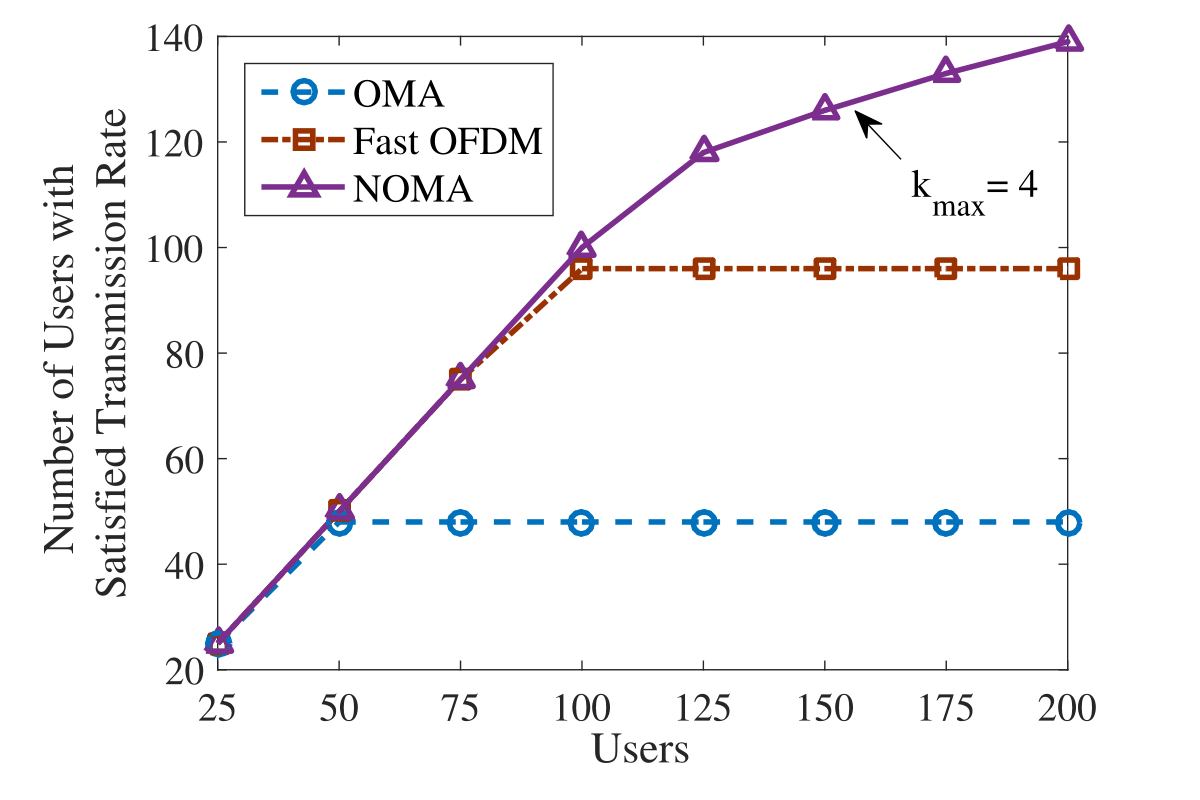}
	\caption{The comparison between NOMA, OMA and fast OFDM in terms of the number of users with satisfied rate requirements.}
	\label{fig:ConnectedUsers1}
\end{figure}

Fig.~{\ref{fig:ConnectedUsers1}} compares the performance of the proposed NOMA based NB-IoT with the OMA and the fast OFDM approaches with respect to the number of the MTC devices with satisfied rate requirements. Note that, the OMA scheme cannot support more than 48 users as it allocates each sub-carrier of an NB-IoT system to only one user. NOMA outperforms both the fast OFDM and the OFDMA (as an OMA technique), and facilitates a higher number of successfully connected MTC devices.

\section{Conclusion}\label{sec:conclusion}
In this paper, we have proposed a power domain NOMA scheme with user clustering in an NB-IoT system. In particular, the MTC devices are assigned to different ranks within the NOMA clusters where they transmit over the same frequency resources. Then, we have formulated an optimization problem to maximize the total throughput of the network by optimizing the resource allocation of MTC devices and NOMA clustering while satisfying the transmission power and QoS requirements. We have further designed an efficient heuristic algorithm to solve the proposed optimization problem by jointly optimizing NOMA clustering and resource allocation of MTC devices. Finally, we have presented simulation results to validate the efficiency of our proposal.

\bibliography{NOMAAidedNarrowbandIoT}
\bibliographystyle{IEEEtr}

\end{document}